\documentstyle[aps,multicol,epsfig]{revtex}
\begin{document}
\title{Squeezing of electromagnetic field in a cavity by electrons in Trojan states}
\author{Piotr Kocha\'nski\footnote{email: piokoch@cft.edu.pl}$^{1,2}$, Zofia Bialynicka-Birula$^{2,3}$
and Iwo Bialynicki-Birula\footnote{email: birula@cft.edu.pl}$^{1,2,4}$}
\address{Center for Theoretical Physics$^1$, College of Science$^2$,
Institute of Physics$^3$\\ Al. Lotnik\'ow 32/46, 02-668 Warsaw, Poland \\and
Institute of Theoretical Physics, Warsaw University$^4$}
 \maketitle
\begin{abstract}
The notion of the Trojan state of a Rydberg electron, introduced by
I.Bialynicki-Birula, M.Kali\'nski, and J.H.Eberly (Phys. Rev. Lett. {\bf 73},
1777 (1994)) is extended to the case of the electromagnetic field quantized in a
cavity. The shape of the electronic wave packet describing the Trojan state is
practically the same as in the previously studied externally driven system. The
fluctuations of the quantized electromagnetic field around its classical value
exhibit strong squeezing. The emergence of Trojan states in the cylindrically
symmetrical system is attributed to spontaneous symmetry braking.
\end{abstract}
\pacs{PACS numbers: 32.80.-t, 42.50.-p, 42.50.Ct}
\begin{multicols}{2}

\section{Introduction}

The possibility of creating stationary, nondispersive, localized, wave packets
describing a highly excited electron in a hydrogen atom strongly driven by
circularly polarized microwave radiation was predicted theoretically several
years ago \cite{ibb+mk+jhe94} and confirmed in numerous publications
\cite{uzer95,kebb,zakrz} (for recent reviews of the subject see
\cite{uzer97a,uzer97b,uzer00}). Such electronic states are called Trojan wave
packets by analogy with the cloud of Trojan asteroid in the Sun-Jupiter system.

In all previous studies the microwave field was treated as an external,
classical wave. Dressing of an electron by such a wave of a suitably chosen
intensity and the frequency equal to the Kepler frequency of the electron on the
Rydberg orbit makes the Trojan wave packets highly stable. Their life-time is of
the order of one second \cite{zbb_ibb,dz}, which makes them an interesting
object of study for theoretical and perhaps even for practical reasons.

In the present paper a similar problem of nondispersive electronic wave packets
is studied for an atom interacting with the {\em quantized} electromagnetic
field. Such an approach allows for a fully dynamical treatment of an autonomous
atom-field system. It automatically includes a back reaction of the atom on the
electromagnetic field. Thus, one can study both the dynamics and the statistical
properties of the electromagnetic radiation. Our study fully confirms the
existence of Trojan states of the Rydberg electron in this new regime with
almost exactly the same shape of the wave packet. The back reaction of the
electron on the electromagnetic field pushes the field frequency off resonance.
The quantum fluctuations of the electromagnetic field exhibit strong squeezing.

\section{Hydrogen atom in a cavity}

Anticipating the role of highly populated, discrete modes of the microwave field
in the formation of Trojan electronic states, we consider a hydrogen atom in a
microwave cavity. In the presence of a cavity we can separate a finite number of
relevant degrees of freedom whereas in free space we would have to deal with a
continuous spectrum which precludes the existence of localized stationary states
of the system.

To allow for the rotational symmetry of the atom-field system we choose a
cylindrical cavity. Its dimensions will be large enough to justify the dipole
approximation in the coupling of hydrogen atom with the lowest cavity modes. The
atom placed in the middle of cavity interacts only with ${\rm TE}_{1n}$ modes.
For definitness we choose the two (degenerate) lowest modes of this type ($n=1$)
(labeled by $X$ and $Y$) for which the mode functions have the form
\begin{mathletters}
\begin{equation}
{\mathbf{E}}^{X} = i{\cal N}\omega\sin\frac{\pi z}{L}
\,{\mathbf{e}}_{z}\times\nabla_{\bot}{\rm J}_{1}(x_{11}r/R)\sin\varphi,
\end{equation}
\begin{equation}
{\mathbf{B}}^{X} = -\frac{{\cal N}\pi}{L}\cos\frac{\pi z}{L} \nabla_{\bot}{\rm
J}_{1}(x_{11}r/R)\sin\varphi,
\end{equation}
\begin{equation}
{\mathbf{E}}^{Y} = -i{\cal N}\omega\sin\frac{\pi z}{L}
\,{\mathbf{e}}_{z}\times\nabla_{\bot}{\rm J}_{1}(x_{11}r/R)\cos\varphi,
\end{equation}
\begin{equation}
{\mathbf{B}}^{Y} = \frac{{\cal N}\pi}{L}\cos\frac{\pi z}{L} \nabla_{\bot}{\rm
J}_{1}(x_{11}r/R)\cos\varphi,
\end{equation}
\end{mathletters}
where $R$ and $L$ are the radius and the length of the cavity and $x_{11}$ is the
first (the smallest) solution of the equation $d{\rm J}_{1}(x)/dx = 0$. The
$z$-axis is taken along the cylinder axis and $\nabla_{\bot} =
(\partial/\partial x,\partial/\partial y)$. The frequency of the modes and the
effective wave vector are given as
\begin{eqnarray}
\omega &=& \frac{c}{R}(x_{11}^2+(\pi R/L)^2)^{\frac{1}{2}},\\
k &=& \sqrt{(\omega/c)^2+(\pi/L)^2}.
\end{eqnarray}
The value of the normalization constant ${\cal N}$ has been obtained in Ref.\
\cite{mar+plk96},
\begin{equation}
{\cal N} = \frac{x_{11}}{k^2R^2}\sqrt{\frac{\hbar} {2\pi\epsilon_{0}L
\omega(1-1/x_{11}^2) {\rm J}_{1}^{2}(x_{11})}},
\end{equation}
from the requirement that the energy per one photon in a mode is equal to $\hbar
\omega$.

At the position ${\mathbf{r}}_{A}=(0,0,L/2)$ of the center of the atom the
orthogonal field vectors ${\mathbf{E}}^{X}$ and ${\mathbf{E}}^{Y}$ point
respectively in $x$ and $y$ direction and are given by simple formulas:
\begin{mathletters}
\begin{eqnarray}
{\mathbf{E}}^{X}(\mathbf{r}_{A}) &=& \frac{-i{\cal
N}\omega x_{11} }{2R}\,\, {\mathbf{e}}_{x},\\
{\mathbf{E}}^{Y}(\mathbf{r}_{A}) &=& \frac{-i{\cal
N}\omega x_{11} }{2R}\,\, {\mathbf{e}}_{y},\\
{\mathbf{B}}^{X}({\mathbf{r}}_{A}) &=&  0,\;\;
 {\mathbf{B}}^{Y}({\mathbf{r}}_{A}) = 0.
\end{eqnarray}
\end{mathletters}
The relevant part of the electric and magnetic field in the cavity can be
written in the form
\begin{mathletters}
\begin{eqnarray}
{\mathbf{E}} &=& {\mathbf{E}}^{X}a_{X}+{\mathbf{E}}^{X\ast}a^{\ast}_{X}+
{\mathbf{E}}^{Y} a_{Y}+{\mathbf{E}}^{Y\ast}a^{\ast}_{Y}\,,\\
{\mathbf{B}} &=& {\mathbf{B}}^{X}a_{X}+{\mathbf{B}}^{X\ast}a^{\ast}_{X}+
{\mathbf{B}}^{Y} a_{Y}+{\mathbf{B}}^{Y\ast} a^{\ast}_{Y}\,,
\end{eqnarray}
\end{mathletters}
where $a_{X}$ and $a_{Y}$ are the dimensionless mode expansion amplitudes.

In the laboratory frame the dynamics of the atom-field system is governed by the
Hamiltonian
\begin{eqnarray}
 H_{L} = \frac{{\mathbf{p}}^2}{2m}-\frac{e^2}{4\pi\epsilon_{0}r}
 &-& e{\mathbf{r}}\!\cdot\!{\mathbf{E}}({\mathbf{r}}_{A})\nonumber\\
 &+& \frac{1}{2}\int(\epsilon_0{\mathbf{E}}^2+{\mathbf{B}}^2/\mu_0)d^{3}{\mathbf{r}},
\end{eqnarray}
where ${\mathbf{r}} = (x,y,z)$ is the position of the electron relative to the
center of the atom ${\mathbf{r}}_{A}$. The Hamiltonian $H_L$ describes the
mutual interaction of the atomic electron with the chosen cavity modes. We can
rewrite $H_{L}$ using the amplitudes $a_{X}$ and $a_{Y}$, or more conveniently,
using their real combinations:
\begin{mathletters}
\begin{eqnarray}
P_{x} &=& \frac{-i}{\sqrt{2}}(a_{X}-a^{\ast}_{X}),\:\: P_{y} =
\frac{-i}{\sqrt{2}}(a_{Y}-a^{\ast}_{Y}),
\end{eqnarray}
\begin{eqnarray}
Q_{x} = \frac{1}{\sqrt{2}}(a_{X}+a^{\ast}_{X}),\:\:
 Q_{y} = \frac{1}{\sqrt{2}}(a_{Y}+a^{\ast}_{Y}),
\end{eqnarray}
\end{mathletters}
where the dimensionless vectors ${\mathbf{P}}$ and ${\mathbf{Q}}$ represent the
electric field and the magnetic induction,
\begin{eqnarray}\label{ham0}
 H_{L} = \frac{{\mathbf{p}}^2}{2m} -\frac{e^2}{4\pi\epsilon_{0}r}
 - e{\cal E}{\mathbf{r}}\!\cdot\!{\mathbf{P}}
 + \frac{\hbar\omega}{2}({\mathbf{P}}^2+{\mathbf{Q}}^2).
\end{eqnarray}
The field amplitude ${\cal E}$ is
\begin{equation}
{\cal E} = \frac{{\cal N}\omega x_{11}}{R\sqrt{2}}.
\end{equation}
We have found it convenient to use the natural units for our problem derived
from the field frequency for the energy, length, and momentum: $\hbar\omega,
\sqrt{\hbar/m\omega}, \sqrt{\hbar m\omega}$. The Hamiltonian (\ref{ham0}) in
these units takes on the following form
\begin{eqnarray}\label{ham1}
\nonumber H_{L} = \frac{{\mathbf{p}}^2}{2} -\frac{\tilde{q}}{r} -
\gamma\,{\mathbf{r}}\!\cdot\!{\mathbf{P}} +
\frac{{\mathbf{P}}^2+{\mathbf{Q}}^2}{2},
\end{eqnarray}
where the dimensionless parameters $\tilde{q}$ and $\gamma$ characterizing the
strength of the Coulomb field and the atom-field coupling are
\begin{eqnarray}
 \tilde{q} =
 \frac{e^2}{4\pi\epsilon_{0}\hbar\omega}\sqrt{\frac{m\omega}{\hbar}},\;\;
 \gamma = \frac{e{\cal E}}{\hbar \omega}\sqrt{\frac{\hbar}{m\omega}}.
\end{eqnarray}

\section{Classical solutions}

We are interested in special solutions corresponding to the Trojan states in the
external electromagnetic wave introduced in Ref.\,\cite{ibb+mk+jhe94}. Since
these states  describe electronic wave packets rotating around the nucleus along
circular orbits, we transform the Hamiltonian $H_{L}$ to the frame rotating
around the $z$-axis with the angular velocity $\Omega$. The transformed
Hamiltonian is
\begin{eqnarray}\label{ham}
H = \frac{{\mathbf{p}}^2}{2} -\frac{\tilde{q}}{r} -
\gamma{\mathbf{r}}\!\cdot\!{\mathbf{P}} +\frac{{\mathbf{P}}^2+{\mathbf{Q}}^2}{2}
-\kappa(M_{z}^{A}+M_{z}^{F}),
\end{eqnarray}
where $\kappa = \Omega/\omega$. The $z$-components of the angular momenta of the
electron and of the electromagnetic field are $M_{z}^{A} = xp_{y}-yp_{x}$ and
$M_{z}^{F} = (Q_{x}P_{y}-Q_{y}P_{x})$. In this frame, the rotational states will
appear as stationary states of the Hamiltonian. We would like to stress that the
Hamiltonian $H$ cannot be identified with the energy because of the appearance
of the inertial forces in the rotating frame.

\begin{figure}
\epsfxsize=8.0cm \epsffile{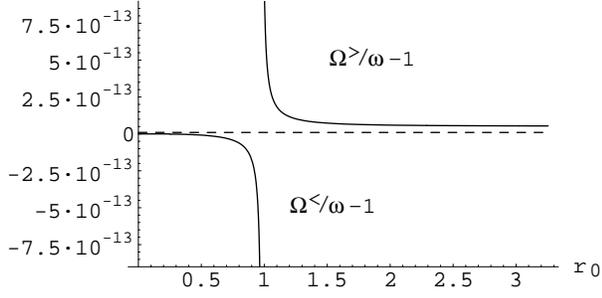} \caption{Two branches of the frequency of
rotation $\Omega(r_0)$. The upper curve corresponds to the Trojan states and the
lower curve corresponds to the anti-Trojan states;
$r_0$ is measured in units 
of $3600 a_0$ ($a_0$ is the atom Bohr radius). } 
\label{omegas}
\end{figure}

\begin{figure}
\epsfxsize=8.0cm \epsffile{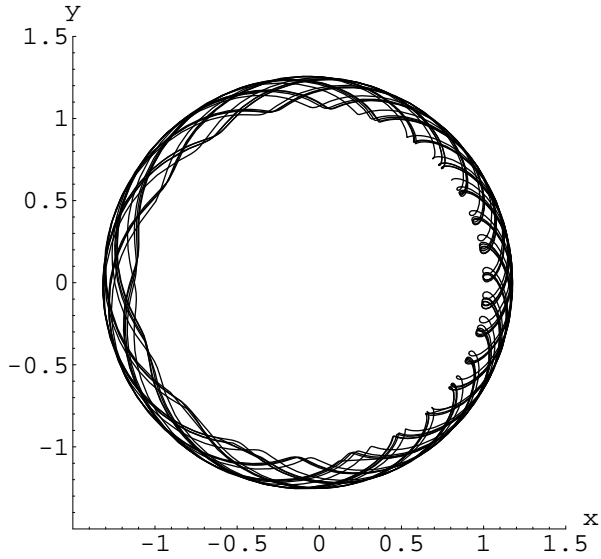} \caption{A typical classical
trajectory in the rotating frame near the anti-Trojan equilibrium position. The
motion extends almost uniformly over the whole circle but the electron spends a
little more time in the right half of the circle.} \label{anti}
\end{figure}

To emphasize the rotational symmetry of our problem we introduce following
variables for the electromagnetic field
\begin{mathletters}
\begin{eqnarray}
Q_{+}=\frac{Q_x - P_y}{\sqrt{2}},\:\: Q_{-}=\frac{Q_x + P_y}{\sqrt{2}},
\end{eqnarray}
\begin{eqnarray}
P_{+}=\frac{Q_y + P_x}{\sqrt{2}},\:\: P_{-}=\frac{P_x - Q_y}{\sqrt{2}},
\end{eqnarray}
\end{mathletters}
corresponding to the left and right circular polarization. In terms of these
variables the Hamiltonian takes on the form
\begin{eqnarray}
\label{hrot} \nonumber H &=& \frac{\mathbf{p}^2}{2}-\frac{\tilde{q}}{r} -
\frac{\gamma}{\sqrt{2}}\bigl(x(P_{+}+P_{-})+y(Q_{-}-Q_{+})\bigr) \\
 &+&\frac{1+\kappa}{2}\left(P_{+}^{2}+Q_{+}^{2}\right)+
\frac{1-\kappa}{2}\left(P_{-}^{2}+Q_{-}^{2}\right) - \kappa M_{z}^{A}.
\nonumber\\
&~&
\end{eqnarray}
The kinetic part of the field Hamiltonian is made up of two terms: co-rotating
and counter-rotating. Linear stability analysis shows that both parts are
necessary for the existence of the nontrivial equilibrium solution. From the
Hamiltonian (\ref{ham}) we derive the evolution equations:
\begin{mathletters}
\label{evolution}
\begin{equation}
\dot{x} = p_{x}+\kappa y,
\end{equation}
\begin{equation}
\dot{y} = p_{y}-\kappa x,
\end{equation}
\begin{equation}
\dot{z} = p_{z},
\end{equation}
\begin{equation}
\dot{Q}_{+} = (1+\kappa)P_{+} - \gamma x/\sqrt{2},
\end{equation}
\begin{equation}
\dot{Q}_{-} = (1-\kappa)P_{-} - \gamma x/\sqrt{2},
\end{equation}
\begin{equation}
\label{eq_cond1} \dot{p}_{x} =
 -\frac{\tilde{q}x}{r^{3}} + \frac{\gamma
(P_{+}+P_{-})}{\sqrt{2}} + \kappa p_{y},
\end{equation}
\begin{equation}
\label{eq_cond2} \dot{p}_{y} =
 -\frac{\tilde{q}y}{r^{3}} + \frac{\gamma
(Q_{-}-Q_{+})}{\sqrt{2}}-\kappa p_{x},
\end{equation}
\begin{equation}
\dot{p}_{z} = -\frac{\tilde{q}z}{r^{3}},
\end{equation}
\begin{equation}
\dot{P}_{+} = -(1+\kappa) Q_{+} - \gamma y/\sqrt{2},
\end{equation}
\begin{equation}
\dot{P}_{-} = -(1-\kappa) Q_{-} + \gamma y/\sqrt{2}.
\end{equation}
\end{mathletters}
The time-independent solutions of these equations describe the stationary states
of our system. Equating the left hand side of Eqs.\,(\ref{evolution}) to zero, we
obtain
\begin{mathletters}
\label{eqsol}
\begin{equation}
x^{eq} = r_{0}\cos\varphi,\;\;  y^{eq} = r_{0}\sin\varphi,\;\;z^{eq}=0,
\end{equation}
\begin{equation}
p_{x}^{eq} = -\kappa r_{0}\sin\varphi, \;\; p_{y}^{eq} = \kappa r_{0}\cos\varphi,
\;\; p_{z}^{eq} = 0,
\end{equation}
\begin{equation}
Q_{+}^{eq} = \frac{-\gamma r_{0}\sin\varphi}{(\kappa+1)\sqrt{2}}, \;\; Q_{-}^{eq}
= \frac{-\gamma r_{0}\sin\varphi}{(\kappa-1)\sqrt{2}},
\end{equation}
\begin{equation}
P_{+}^{eq} = \frac{\gamma r_{0}\cos\varphi}{(\kappa+1)\sqrt{2}},\;\; P_{-}^{eq} =
\frac{-\gamma r_{0}\cos\varphi}{(\kappa-1)\sqrt{2}}.
\end{equation}
\end{mathletters}
In addition, the equations (\ref{eq_cond1}) and (\ref{eq_cond2}) give the
equilibrium condition:
\begin{eqnarray}
\label{stabcond}\frac{\tilde{q}}{r_0^3} = \kappa^2 - \frac{\gamma^2}{\kappa^2-1}.
\end{eqnarray}
This equilibrium condition can be used to express the equilibrium radius $r_0$
in terms of the frequency of the cavity mode $\omega$ and the frequency of
rotation $\Omega$
\begin{eqnarray}
 r_0(\Omega) = \tilde{q}^{1/3}
 \left((\Omega/\omega)^2-\frac{\gamma^2}{(\Omega/\omega)^2-1}\right)^{-1/3},
\end{eqnarray}
or, alternatively, to express the frequency of rotation in terms of $\omega$ and
$r_0$. The equilibrium condition (\ref{stabcond}) has two solutions for $\Omega$,
denoted by $\Omega^>(r_0)$ and $\Omega^<(r_0)$,
\begin{mathletters}
\begin{eqnarray}
\Omega^>(r_0) = \frac{\omega}{\sqrt{2}}\sqrt{1+\tilde{q}/r_0^3 +
\sqrt{(1-\tilde{q}/r_0^3)^2+4\gamma^2}},\\
\Omega^<(r_0) = \frac{\omega}{\sqrt{2}}\sqrt{1+\tilde{q}/r_0^3 -
\sqrt{(1-\tilde{q}/r_0^3)^2+4\gamma^2}}.
\end{eqnarray}
\end{mathletters}
Both solutions exist for all values of $r_0$. The solution $\Omega^>(r_0)\,$
($\Omega^>(r_0)$) gives the frequency that is always higher (lower) than the
cavity frequency $\omega$ (see Fig.\ \ref{omegas}). The higher frequency (larger
centrifugal force) requires the electric field to be directed towards the
nucleus, whereas the lower frequency requires the field pointing outwards. The
first case corresponds to the Trojan states whereas the second to the so called
anti-Trojan states. In the previous study \cite{kal_eb}, when the electromagnetic
field has been treated as a given external wave, the anti-Trojan states were
found to be classically unstable. The classical stability obtained in the
present study is due to the detuning from the exact resonance. Since the
localization of the electronic wave packet is much worse for the anti-Trojan
states (classically, the trajectories in the rotating frame are spread almost
evenly around the whole circle, cf. Fig.\ \ref{anti}), we will restrict ourselves
to the Trojan states only. Hence, in what follows, we shall only consider the
solution $\Omega^>(r_0)$.

Note, that for $\omega = \Omega$ (i.e. $\kappa= 1$), we have only a trivial
result ${\mathbf{r}}^{eq} = {\mathbf{p}}^{eq} =
{\mathbf{P}}^{eq}={\mathbf{Q}}^{eq} = 0$ which in the classical model of an
atom, means that ``the electron has fallen onto the nucleus'' and the electric
field is zero. Thus, every nontrivial solution requires the presence of a
detuning ($\omega\neq\Omega$) between the cavity frequency and the Kepler
frequency. This phenomenon is known as the frequency pushing and is a direct
consequence of the mutual atom-field interaction. This detuning has been absent
in all previous approaches where the atom was driven by an external wave.

The equations (\ref{evolution}) have a continuum of time-independent solutions
that can be labeled by $r_0$ and the angle $\varphi$ in the $x$-$y$ plane. These
solutions describe in the laboratory frame a classical electron circulating
around the nucleus at the distance $r_0$. The orbit of the electron is confined
to the $x$-$y$ plane. The electron is dressed by the classical electromagnetic
field
\begin{equation}
{\mathbf{E}}= -\frac{{\cal E}\sqrt{2}}{\kappa^2-1}\;\frac{e{\cal E}r_0}
{\hbar\omega} \;(\sin\varphi \; \mathbf{e}_{y}+\cos\varphi \; \mathbf{e}_{x}),
\end{equation}
which has a resonance dependence on the parameter $\kappa$. Note that the
electric field changes its sign when the frequency of rotation passes through
the resonance.

\begin{figure}
\epsfxsize=8.0cm  \leavevmode \epsffile{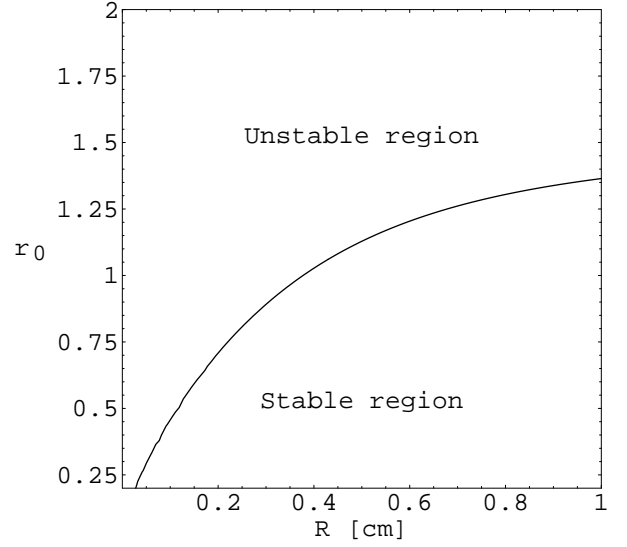} \caption{The boundary
between the stable and unstable regions of classical equilibrium;
$r_0$ is measured in units 
of $3600 a_0$ ($a_0$ is the atom Bohr radius).
}
\label{stability}
\end{figure}

Next, we expand the Hamiltonian around a time-independent solution and
investigate its linear stability. The motion will be stable if all
eigenfrequencies are real. The characteristic equation for this problem has the
form
\end{multicols}
\begin{eqnarray}
&&\lambda^2(\lambda^2-q_r)\Big(\lambda^6 - (4+q_r+2)\Gamma)\lambda^4\nonumber\\
 &&+ (5-3/5q_r+q_r^2/2+4\gamma^2+(4+5/2q_r\Gamma)\lambda^2\nonumber\\
 &&- (2+5/2q_r-5q_r^2+q_r^3+8\gamma^2+14q_r\gamma^2
 +(2+7/2q_r+q_r^2/2+8\gamma^2)\Gamma))\Big) = 0,
\end{eqnarray}
\begin{multicols}{2}
\noindent where $q_r = \tilde{q}/r_0^3$ and $\Gamma =
\sqrt{1-2q+q^2+4\gamma^2}$. The first ($\lambda=0$) frequency in our problem
corresponds to the rotation of the whole system and it is a reflection of the
rotational symmetry. The second frequency ($\lambda=\sqrt{\tilde{q}/r_0^3}$)
corresponds to the motion in the $z$-direction that (in the linear
approximation) is decoupled from the motion in the $x$-$y$ plane. The remaining
3 frequencies correspond to the motion of the electron coupled to the
electromagnetic field. We shall not produce the analytical expressions for these
eigenfrequencies but in Fig.\ \ref{stability} we plot the region of stability in
the $R$-$r_0$ plane.

\begin{figure}
\epsfxsize=8.0cm \epsffile{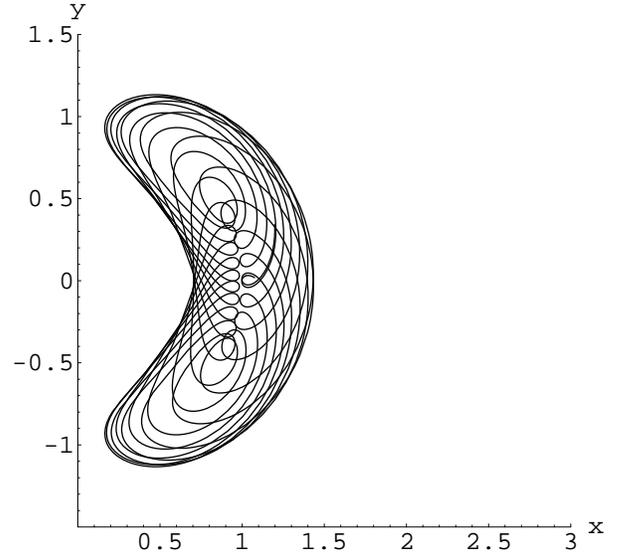} \caption{Classical electron trajectory
projected into the $x$-$y$ plane; $x$ and $y$ are measured in units of $r_0$. The
trajectory started at time $t=0$ from the equilibrium position (\ref{eqsol})
with the initial momenta ${\bf p} = m\omega r_0(0.02,\kappa+0.07,0.02)$ and is
plotted for the time interval $(1400\,T, 1500\,T)$.} \label{xymotion}
\end{figure}

\begin{figure}
\epsfxsize=8.0cm \epsffile{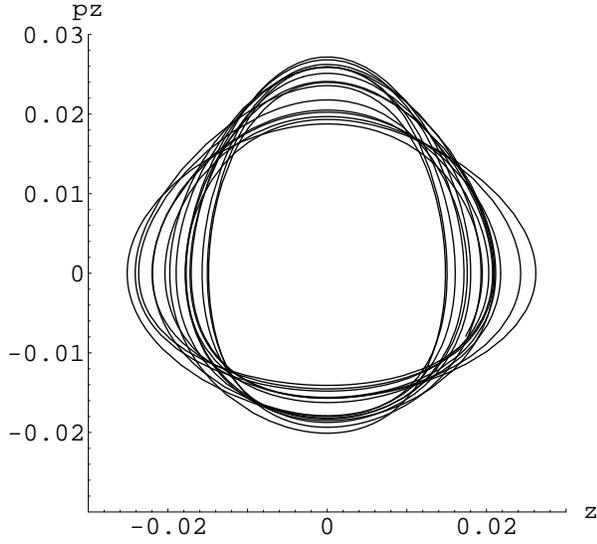} \caption{Classical motion of the
electron projected into the phase space $z$-$p_z$; $z$ is measured in units of
$r_0$ and $p_{z}$ is measured in units of $m\omega r_0$. The trajectory started
at time $t=0$ from the equilibrium position (\ref{eqsol}) with the initial
momenta ${\bf p} = m\omega r_0(0.02,\kappa+0.07,0.02)$ and is plotted for the
time interval $(1400\,T, 1500\,T)$.} \label{zphase}
\end{figure}

The stability can also be studied numerically and the calculations of the
classical trajectories fully confirm the stability of the equilibrium solution.
In Fig.\ \ref{xymotion} we plotted the projection of a typical electron
trajectory on the $x$-$y$ plane for the time interval $(1400\,T, 1500\,T)$,
where $T=1/\omega$. The trajectory started at the equilibrium position ${\bf r} =
r_0(1,0,0)$ with the initial momenta ${\bf p} = m\omega
r_0(0.02,\kappa+0.07,0.02)$. As we see, the electron follows a rather
complicated, but bounded, trajectory. Obviously, if we choose $p(t=0)$
sufficiently large the electron will eventually leave the vicinity of the
equilibrium point. In Fig.\ \ref{zphase} we show the $z$-$p_z$ cross-section of
the phase space for the same trajectory. This phase-space trajectory resembles
the trajectory of a simple harmonic oscillator. Indeed, as we have seen, in the
linearized evolution equations, the motion in the $z$-direction is purely
harmonic. Thus, the interesting dynamics of the electron is found in the motion
confined to the $x$-$y$ plane and in what follows we shall treat our problem as
two-dimensional.

\begin{figure}
\epsfxsize=8.0cm  \leavevmode \epsffile{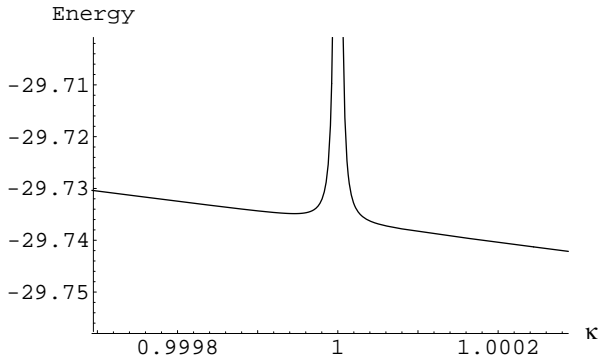} \caption{The energy
$E(\kappa)$ plotted in units of $\hbar\omega$.} \label{energy}
\end{figure}

Since our system is conservative, it has a well defined energy $H_L$. We have
calculated its value $E(\kappa)$ for all those solutions that in the rotating
frame are determined by Eqs.\,(\ref{eqsol}). This energy is given by the formula
\begin{eqnarray}
\label{en_form} E(\kappa) = \frac{m\omega^2r_{0}^2(\kappa)}{2}
\left(\frac{\gamma^2}{(\kappa^2-1)^2}(5\kappa^2-3)-\kappa^2\right).
\end{eqnarray}
and is plotted in Fig.\ \ref{energy} as a function of $\kappa$. The infinite
growth of the energy near the resonance ($\kappa = 1)$ expresses the phenomenon
of the frequency pushing.

\section{Quantum effects}
In order to study the quantum effects for the electron as well as for the
electromagnetic field we will apply the procedure of the quantization around the
classical solution (\ref{eqsol}). A similar quantization method has been used
before, for example in nonlinear optics to describe quantum fluctuations around
the classical solitons in fibers \cite{lai+haus}. Here, the quantization will
lead to the description of the electron in terms of a quantum mechanical wave
packet orbiting along the classical trajectory and, at the same time, will
reveal quantum fluctuations of the electromagnetic field around its classical
value.

As a starting point we choose the Hamiltonian (\ref{hrot}) in which all
variables are treated as operators and we express them as sums of their classical
parts and the quantum corrections $\hat{{\mathbf{r}}} =
{\mathbf{r}}^{eq}+{\mathbf{r}}$,
${\hat{{\mathbf{p}}}=\mathbf{p}}^{eq}+{\mathbf{p}}$,
${\hat{{\mathbf{Q}}}=\mathbf{Q}}^{eq}+{\mathbf{Q}}$,
${\hat{{\mathbf{P}}}=\mathbf{P}}^{eq}+{\mathbf{P}}$. The classical parts
represent equilibrium solutions (\ref{eqsol}) found in the preceding Section. In
order to simplify the notation, we have not attached any labels to the operators
of quantum corrections $(\mathbf{r},\mathbf{p},\mathbf{Q},\mathbf{P})$. Next, we
expand the Hamiltonian around the classical equilibrium solution neglecting all
terms higher then quadratic in the quantum corrections. To proceed along these
lines, we have to choose one solution, labeled by $\varphi_0$, from the whole
family of equilibrium solutions. Making this choice we break the rotational
symmetry of the Hamiltonian.

This mechanism of selection of a specific classical solution resembles the
spontaneous symmetry breaking. Spontaneous symmetry breaking is present in many
branches of physics. It explains the appearance of deformed nuclei, the
formation of magnetic domains in ferromagnetic materials, or the emergence of
Higgs particles in the Glashow-Weinberg-Salam model of electroweak interactions.
In all these cases the symmetry is broken by the choice of a particular ground
state. In our case, however, we do not break the symmetry by choosing a ground
state but by choosing an equilibrium state of the Hamiltonian that is very far
from the ground state of the system.

Once we have chosen some $\varphi_0$, we can rotate the frame of reference, so
that the direction given by $\varphi_0$ is along the $x$ axis. The quadratic
Hamiltonian is
\begin{eqnarray}
\label{qham} \nonumber H_{Q}&=&\frac{\mathbf{p}^2}{2} -
\frac{\gamma}{\sqrt{2}}[x(P_{+}+P_{-})+y(Q_{-}-Q_{+})] \\
\nonumber &+&\frac{1+\kappa}{2}\left(P_{+}^{2}+Q_{+}^{2}\right) +
\frac{1-\kappa}{2}\left(P_{-}^{2}+Q_{-}^{2}\right) \\
&-&q\kappa^2x^2+\frac{q\kappa^2y^2}{2}+\frac{q\kappa^2z^2}{2} -\kappa(x p_y - x
p_x),
\end{eqnarray}
where the parameter $q$, the ratio of the Coulomb force to the centrifugal force,
\begin{eqnarray}
 q = \frac{e^2}{4\pi\epsilon_{0}mr_0^3\Omega^2} = \frac{\tilde{q}}{r_0^3\kappa^2},
\end{eqnarray}
has been introduced to achieve the full correspondence with the notation used
before \cite{ibb+mk+jhe94,ibb_zbb} in the description of Trojan states. Note,
that in this Hamiltonian the quadratic term $q\kappa^2x^2$ enters with the
negative coefficient. If it were not for the rotational term, such a Hamiltonian
would not have any stable points. In our case, however, the stability can be
achieved for a particular choice of $\gamma$, $q$ and $\kappa$ .

We look for the fundamental solution of the Schr\"odinger equation with the
Hamiltonian $H_Q$ in the form of a four-dimensional Gaussian function
\begin{equation}\label{gaussian}
\psi =
N\exp{\left(-\frac{1}{2}\,{\mathbf{X}}\!\cdot\!A\!\cdot\!{\mathbf{X}}\right)},
\end{equation}
${\mathbf{X}}=(x, y, Q_{+}, Q_{-})$, and
\begin{eqnarray}
A= \left(\begin{array}{cccc}
a_{11} & ia_{12} & ia_{13} & ia_{14} \\
ia_{12} & a_{22} & a_{23} & a_{24} \\
ia_{13} & a_{23} & a_{33} & a_{34} \\
ia_{14} & a_{23} & a_{34} & a_{44}
\end{array}\right)
\end{eqnarray}
Inserting this ansatz into the Schr\"odinger equation we obtain 10 algebraic,
nonlinear equations for the parameters $a_{ij}$:
\end{multicols}
\begin{mathletters}
\begin{equation}
 -2\kappa^2q - a_{11}^2 + 2 \kappa  a_{12} + a_{12}^2 - 2 \gamma  a_{13} +
 a_{13}^2(1 + \kappa)- 2\gamma a_{14} + a_{14}^2(1 - \kappa)=  0,
\end{equation}
\begin{equation}
 a_{11} a_{12} + a_{12} a_{22} - \gamma  a_{23} + a_{13} a_{23} -
 \gamma  a_{24} + a_{14} a_{24} + \kappa  \left( -a_{11} + a_{22} +
 a_{13} a_{23} - a_{14} a_{24} \right) = 0,
\end{equation}
\begin{equation}
 a_{11} a_{13} + a_{12} a_{23} - \gamma  a_{33} + a_{13} a_{33} - \gamma  a_{34} +
 a_{14} a_{34} + \kappa  \left( a_{23} + a_{13} a_{33} - a_{14} a_{34} \right) = 0,
\end{equation}
\begin{equation}
 a_{11} a_{14} + a_{12} a_{24} - \gamma  a_{34} + a_{13} a_{34} - \gamma  a_{44} +
 a_{14} a_{44} + \kappa  \left( a_{24} + a_{13} a_{34} - a_{14} a_{44} \right) = 0,
\end{equation}
\begin{equation}
 \kappa^2q - 2\kappa a_{12} + a_{12}^2 - a_{22}^2 - a_{23}^2(1 + \kappa) - a_{24}^2(1- \kappa) = 0,
\end{equation}
\begin{equation}
 -\gamma - \kappa  a_{13} + a_{12} a_{13} - a_{22} a_{23} - a_{23} a_{33} -
 \kappa  a_{23} a_{33} - a_{24} a_{34} + \kappa  a_{24} a_{34} = 0,
\end{equation}
\begin{equation}
 \gamma - \kappa  a_{14} + a_{12} a_{14} - a_{22} a_{24} - a_{23} a_{34} -
 \kappa  a_{23} a_{34} - a_{24} a_{44} + \kappa  a_{24} a_{44} = 0,
\end{equation}
\begin{equation}
 1 + \kappa +a_{13}^2 - a_{23}^2 - a_{33}^2(1+\kappa) - a_{34}^2(1-\kappa) = 0,
\end{equation}
\begin{equation}
 a_{13} a_{14} - a_{23} a_{24} - a_{34} \left( \left( 1 + \kappa \right)  a_{33} -
 \left( -1 + \kappa \right)  a_{44} \right) = 0,
\end{equation}
\begin{equation}
 1 - \kappa + a_{14}^2 - a_{24}^2 - a_{44}^2(1- \kappa) - a_{34}^2(1 + \kappa) = 0.
\end{equation}
\end{mathletters}
\begin{multicols}{2}
We can easily solve these equations numerically, but first we want to find a
perturbative solution. In order to do that we write the coupling constant in the
form $\gamma = \bar{\gamma}\sqrt{\kappa-1}$. Obviously $\bar{\gamma} =
\kappa\sqrt{(1-q)(\kappa+1)}$, and we will treat $\bar{\gamma}$ as a small
parameter. Typical values of the parameters are $\kappa = 1.0000001, q =
0.95625$ which give $\bar{\gamma} = 0.06$. One can ask why we can not treat
$\gamma$ (or even simpler, $\kappa -1$) as a perturbation parameter. However, if
we do so we face a problem: the coefficients of the perturbation series are
growing, since they behave as $1/\sqrt{\kappa -1}$. When we tend with $\kappa
-1$ to zero we hit exactly the resonance point and the perturbation expansion
becomes meaningless. On the other hand, when $\bar{\gamma}$ is chosen as an
expansion parameter, all large contributions to the coefficients in the
perturbation expansion $a_{ij}=a^{(0)}_{ij}+\bar{\gamma} a^{(1)}_{ij}+
\bar{\gamma}^2 a^{(2)}_{ij}+\dots$ cancel out.

We calculated analytically the coefficients up to the second order but we
present here the analytic formulas only in the zeroth order and numerical values
of the first and the second order corrections.
\begin{mathletters}
\begin{eqnarray}
a^{(0)}_{11}&=&\kappa \sqrt{\frac{(1 + 2q )(4q - 9q + 8-8s(q))}{9q^2}},\\
 a^{(0)}_{12}&=&\kappa\frac{2 + q - 2s(q)}{3q},\\
 a^{(0)}_{22}&=&\kappa\sqrt{\frac{(1 - q )( 4q - 9q + 8-8s(q))}{9q}},
\end{eqnarray}
\end{mathletters}
where $s(q)=\sqrt{1 + q - 2q}$,
\begin{eqnarray}
a^{(0)}_{13}&=&0,\,\, a^{(0)}_{23}=0,\,\,a^{(0)}_{14}=0,\,\, a^{(0)}_{24}=0,
\nonumber\\
a^{(0)}_{33}&=&1,\,\,a^{(0)}_{34}=0,\,\,a^{(0)}_{44}=1.
\end{eqnarray}
Thus, in the zeroth order, the electronic part of the wave packet is exactly the
same as in the case of externally driven Trojan wave packet \cite{ibb+mk+jhe94}.
The electromagnetic part has a form of a coherent (nonsqueezed) state.

Higher corrections are due to the mutual interaction between the field and the
atom. Numerical values of the parameters $a_{ij}$ are calculated for the cavity
parameters $L=1$ cm, $R=0.32$ cm, which give $\omega=197$ GHz and
$\gamma=3.24\times 10^{-7}$. The detuning $\kappa$ is chosen in such a way, that
the value of $q$ is optimal, $q=0.95625$. As shown in Ref.\,\cite{ibb+mk+jhe94},
the wave packet is then maximally concentrated around the equilibrium point and
its center is located at $r_{0}=3600 a_{0}$ ($a_0$ is the atom Bohr radius). The
expansion coefficients calculated up to the second order are presented in Table
1. In this order we observe the effect of the back reaction of the electron on
the electromagnetic field. However, the coefficients $a_{11}, a_{12}$, and
$a_{22}$ characterizing the shape of the electronic wave packet are the same as
in the zeroth order within the assumed accuracy.

Finally, we present in Table 2 the results of a direct numerical solution of our
set of equations. As we see, almost all the coefficients have been obtained
correctly already in the second order of perturbation theory. Only the $a_{44}$
differs from the exact numerical solution. This can be attributed to the very
slowly convergent perturbation series for this particular coefficient. The
coefficients $a_{13},a_{23},a_{14}$ and $a_{24}$ describing the mixing of the
atomic part of the wave packet with the field part are not zero. Moreover, the
electromagnetic field is strongly affected by the interaction with the atom; the
coefficient $a_{44}$ is significantly different from its value in the zeroth
order.

The four-dimensional Gaussian wave packet (\ref{gaussian}) with the coefficients
$a_{ij}$ calculated numerically describes the fundamental state of the mutually
interacting atom-field system. Owing to its Gaussian form, this state saturates
the multidimensional generalized uncertainty relations for the complete
atom-field system (see \cite{ibb98}). The smallness of the coefficients
$a_{13},a_{23},a_{14}$ and $a_{24}$ expresses the fact that the field and the
atom are only very weakly correlated  in this state. As a result of this, the
saturation of the uncertainty relations is almost exact, separately for both
parts of the wave function. The average values of second moments the electronic
variables calculated with the numerical values of the coefficients taken from
the Table 2 are given in Table 3. This Table exhibits the existence of
correlations between the variables in the $x$ and the $y$ directions. This
requires the use of generalized uncertainty relations for a two-dimensional
system in the form (cf. Ref.\cite{ibb98})

\begin{eqnarray}
\nonumber && \langle x x\rangle \langle p_x p_x\rangle  +
 \langle x y\rangle \langle p_x p_y\rangle \\-
&& \frac{1}{4}( \langle x p_x+p_x x\rangle^2+
\langle x p_y+p_y x\rangle \langle p_x y + y p_x \rangle) \geq \frac{\hbar^2}{4}, \\
\nonumber && \langle y y\rangle \langle p_y p_y\rangle  +
 \langle x y\rangle \langle p_x p_y\rangle \\-
&& \frac{1}{4}(\langle y p_y+p_y y\rangle^2+ \langle y p_x+p_x y\rangle \langle
p_y x + x p_y \rangle) \geq\frac{\hbar^2}{4}.
\end{eqnarray}
Upon substituting the values taken from the Table 3 we find an almost exact
saturation of these relations.

\begin{figure}
\epsfxsize=8.0cm \epsffile{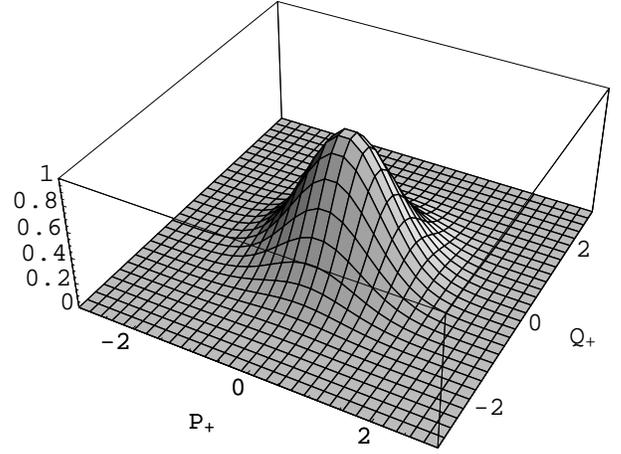} \caption{The $Q_{+}$-$P_{+}$
cross-section of the Wigner function. In this (counter-rotating) mode the
fluctuations of the electromagnetic radiation are not squeezed.} \label{wigplus}
\end{figure}

\begin{figure}
\epsfxsize=8.0cm \epsffile{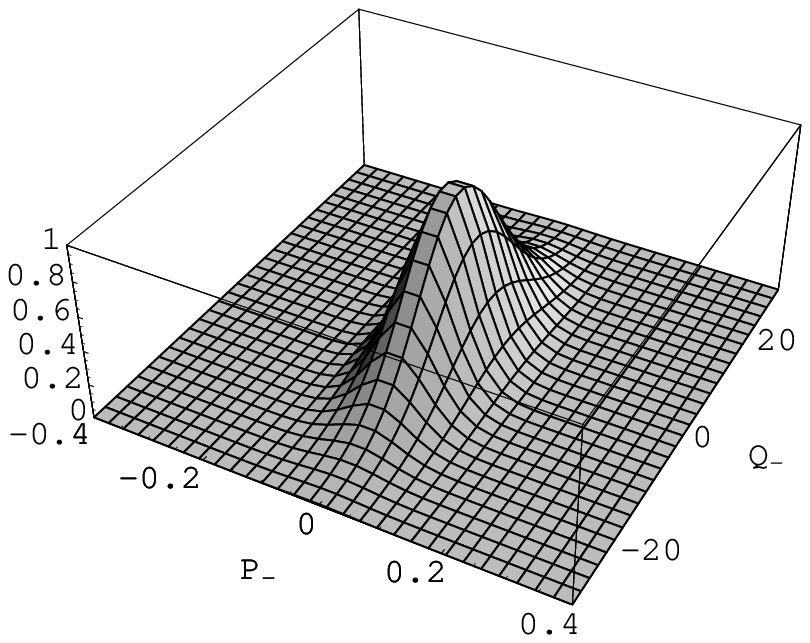} \caption{The $Q_{-}$-$P_{-}$
cross-section of the Wigner function. In this (co-rotating mode) the fluctuations
of the electromagnetic radiation are strongly squeezed. Note the change of the
scale, as compared to Fig.\ \ref{wigplus}.} \label{wigminus}
\end{figure}

Now, we turn to the description of the quantum correlations for the
electromagnetic field in our fundamental state of the atom-field system. The
second moments for the field variables are given in the Table 4. These values of
the correlations imply an almost complete decoupling between the co-rotating and
counter-rotating modes so that the uncertainty is almost saturated separately
for each mode. The state of the field in the counter-rotating mode is a coherent
state but the state in the co-rotating mode is highly squeezed; the ratio of the
correlation for the two quadratures $Q_-$ and $P_-$ is about $3.5\times 10^4$.
However, the fluctuations of the field are still small as compared to the field
value $P^{eq}_- = 1.5\times 10^6$. The plots of the Wigner function in Figs.\
\ref{wigplus} and \ref{wigminus} illustrates the difference in quantum
fluctuations between the counter-rotating and the co-rotating modes.

\section{Discussion}

We have shown that the dynamical treatment of the relevant modes of the
electromagnetic field enables one to reproduce exactly the properties of the
Trojan states studied previously in the presence of a given, external wave.
However, there appear new features totally absent in the previous studies.
First, the back reaction of the electron on the electromagnetic field causes the
detuning from the exact resonance. As a result, the stability region covers now
also the anti-Trojan states, that were before found to be classically unstable.
Second, our analysis has shown that to achieve the equilibrium state of the
mutually interacting atom-field system we must take into account both
polarization modes of the field: co-rotating and counter-rotating. The inclusion
of only the co-rotating mode, as proposed in Ref. \cite{dz}, is not sufficient
to achieve an equilibrium state.

The quantization procedure adopted by us in this work consists in quantizing the
corrections to the classical solution. These quantum corrections describe the
shape of the electronic wave packet and the quantum fluctuations of the
electromagnetic field around its classical value. The field fluctuations turn
out to be significantly different for the two modes: for the counter-rotating
mode the fluctuations are as for the vacuum state whereas for the co-rotating
mode they exhibit strong squeezing.

The choice of one particular solution from the class of equivalent classical
solutions spontaneously breaks the rotational symmetry of the initial
Hamiltonian. The method of quantization around the classical solution used here
can be also applied to a similar problem of electronic Trojan states in a polar
molecule \cite{ibb_zbb}. In this case, the role of the electromagnetic field is
played by the rotating molecular dipole. The application of our method would
require the dynamical treatment of the relevant molecular degrees of freedom.

\section*{Acknowledgments}

We would like to thank Jan Mostowski for fruitful discussions and we acknowledge
the support from the KBN Grant P03B0313.

\begin{table}
\caption{Coefficients characterizing the fundamental state of the atom-field
system calculated up to the second order of perturbation theory.\label{table1}}
\begin{tabular}{lll}
$a_{11}=0.51160 $ & $ a_{12}= 0.78164 $ & $ a_{22}=0.06270 $\\
$a_{33}= 1 $ & $ a_{34}=1.49\times 10^{-12}$  & $ a_{44}=0.50751 $  \\
$a_{13}=7.50\times 10^{-7} $ & $ a_{14}= 4.50\times 10^{-6}$ &  \\
$a_{23}=-5.33\times 10^{-7} $ & $ a_{24}=-7.68\times 10^{-7} $
\end{tabular}
\end{table}

\begin{table}
\caption{Coefficients characterizing the fundamental state of the atom-field
system calculated numerically.\label{table2}}
\begin{tabular}{lll}
$ a_{11}=0.51160 $ & $a_{12}= 0.78164 $& $ a_{22}=0.06270$ \\
$ a_{33}= 1 $ & $ a_{34}=1.40\times 10^{-12} $ & $ a_{44}= 0.00532$  \\
$ a_{13}=7.50\times 10^{-7} $ & $ a_{14}= 4.668\times 10^{-6}$ &  \\
$ a_{23}=-5.33\times 10^{-7} $ & $ a_{24}=-1.34\times 10^{-6} $
\end{tabular}
\end{table}

\begin{table}
\caption{Correlations of positions and momenta for the electronic variables
calculated in the fundamental state of the atom-field system. The position
variables are measured in units of the electron orbit radius $r_0$ and the
momenta are measured in the corresponding unit $m\Omega r_0$.\label{table3}}
\begin{tabular}{ll}
$ \langle x x\rangle = 0.01595 $ & $\langle x p_x+p_x x\rangle = 0$\\
$ \langle y y\rangle = 0.13014 $& $ \langle y p_y+p_y y\rangle = 0 $\\
$ \langle p_x p_x\rangle = 0.08369 $ & $\langle x p_y+p_y x\rangle =-0.02493 $\\
$ \langle p_y p_y\rangle = 0.01026 $ & $\langle y p_x+p_x y\rangle =-0.20345 $\\
$ \langle x y\rangle = 0$ & $\langle p_x p_y\rangle =0  $\\
\end{tabular}
\end{table}

\begin{table}
\caption{Correlations of the electromagnetic field variables calculated in the
fundamental state of the atom-field system.\label{table4}}
\begin{tabular}{ll}
$\langle Q_+ Q_+ \rangle =0.5$  & $\langle (Q_+ P_+ + P_+ Q_+)\rangle = 0$\\
$\langle Q_- Q_- \rangle = 94.05900$ & $\langle (Q_- P_- + P_- Q_-)\rangle = 0$ \\
$\langle P_+ P_+\rangle = 0.5$ & $\langle (Q_+ P_- + P_- Q_+)\rangle =0 $\\
$\langle P_- P_-\rangle = 0.002657$ & $\langle (Q_- P_+ + P_+ Q_-)\rangle =0$  \\
$\langle Q_+ Q_-\rangle = 9.38113\times 10^{-10}$ & $\langle P_+ P_-\rangle
=4.12131\times 10^{-12}$
\end{tabular}
\end{table}

\end{multicols}

\begin{references}
 \bibitem{ibb+mk+jhe94} I. Bialynicki-Birula, M. Kali\'nski and
 J. H. Eberly, Phys. Rev. Lett. {\bf 73}, 1777 (1994).
 An earlier prediction of such states based on a semiclassical approximation
 has been made by H. Klar, Z. Phys. D {\bf 11}, 45 (1989).
 \bibitem{uzer95} D. Farrelly, and T. Uzer, Phys. Rev. Lett. {\bf
 74}, 1720 (1995).
 \bibitem{kebb}  M. Kali\'nski, J. H. Eberly, and I. Bialynicki-Birula, Phys. Rev.
 A {\bf 52}, 2460 (1995).
 \bibitem{zakrz} D. Delande, J. Zakrzewski, and A. Buchleitner,
 Europhys. Lett. {\bf 32}, 107 (1995);
 J. Zakrzewski, D. Delande,  and A. Buchleitner, Phys. Rev. Lett.
 {\bf 75}, 4015 (1995).
 \bibitem{uzer97a}E. Lee, A. F. Brunello, and D. Farrelly,
 Phys. Rev A. {\bf 55}, 2203 (1997).
 \bibitem{uzer97b}C. Cerjan, E. Lee, A. F. Brunello, D. Farrelly, and T.
 Uzer, Phys. Rev. A {\bf 55}, 2222 (1997).
 \bibitem{uzer00} T. Uzer, E. Lee,  D. Farrelly, and A. F. Brunello,
 Cont. Phys. {\bf 41}, 1, (2000).
 \bibitem{kal_eb} M. Kali\'nski and J. H. Eberly, Phys. Rev. Lett. {\bf 77}, 2420
 (1996).
 \bibitem{zbb_ibb} Z. Bialynicka-Birula and I. Bialynicki-Birula, Phys. Rev.
 A {\bf 56}, 1458 (1997).
 \bibitem{dz} D. Delande and J. Zakrzewski, Phys. Rev. A {\bf 58}, 466 (1998).
 \bibitem{mar+plk96} M. A. Rippin and P. L. Knight, J. Mod. Opt {\bf 43}, 807
 (1996).
 \bibitem{uzer96}A. F. Brunello, T. Uzer, and D. Farrelly,
 Phys. Rev. Lett. {\bf 76}, 2874 (1996).
 \bibitem{ibb_zbb} I. Bialynicki-Birula and Z. Bialynicka-Birula, Phys. Rev.
 Lett. {\bf 77}, 4298 (1996).
 \bibitem{lai+haus} H. A. Haus and Y. Lai, J. Opt. Soc. Am. B {\bf
 7}, 386 (1990).
 \bibitem{ibb98} I. Bialynicki-Birula, Acta Phys. Polon. {\bf 29},
 3569 (1998).
\end{references}
\end{document}